\documentclass[a4paper,12pt]{article}
\usepackage{tikz}
\usepackage{tabularx,ragged2e}
\usepackage{pythontex} 
\usetikzlibrary{patterns,perspective}
\usepackage{pgfplots}
   \usetikzlibrary{matrix,decorations.pathreplacing, calc, positioning,fit}
\usetikzlibrary{positioning}
\usetikzlibrary{calc}
\usetikzlibrary{arrows,shapes,backgrounds}
\usepackage{bbding}
\usepackage{bbm}
\usepackage{amsmath}
\usepackage{amsthm}
\usepackage{listings}

\usepackage{bm}
\usepackage{bbold}
\usepackage{appendix}
\usepackage{graphicx}
\usepackage{natbib}
\usepackage{listings}
\usepackage[ruled,vlined]{algorithm2e}
\usepackage{hyperref}
\usepackage{cleveref}
\usepackage{enumitem}
\DeclareMathOperator*{\argmin}{arg\,min}

\DeclareMathAlphabet{\pazocal}{OMS}{zplm}{m}{n}
\newcommand{\unif}{\pazocal{U}}

\newcommand{\bbeta}{{\boldsymbol{\beta}}}

\newcommand{\bX}{\mathbf{X}}
\newcommand{\bx}{\mathbf{x}}
\newcommand{\bb}{\mathbf{b}}
\newcommand{\calR}{\mathcal{R}}
\newcommand{\R}{\mathbbm{R}}

\usepackage[utf8]{inputenc}

\DeclareFixedFont{\ttb}{T1}{txtt}{bx}{n}{12} 
\DeclareFixedFont{\ttm}{T1}{txtt}{m}{n}{12}  

\usepackage{color}
\definecolor{deepblue}{rgb}{0,0,0.5}
\definecolor{deepred}{rgb}{0.6,0,0}
\definecolor{deepgreen}{rgb}{0,0.5,0}

\usepackage{listings}

\newcommand\pythonstyle{\lstset{
		language=Python,
		basicstyle=\ttm,
		morekeywords={self},              
		keywordstyle=\ttb\color{deepblue},
		emph={MyClass,__init__},          
		emphstyle=\ttb\color{deepred},    
		stringstyle=\color{deepgreen},
		frame=tb,                         
		showstringspaces=false
}}

\lstnewenvironment{python}[1][basicstyle=\small]
{
	\pythonstyle
	\lstset{#1}
}
{}

\newcommand\pythoninline[1]{{\pythonstyle\lstinline!#1!}}

\begin{document}
\title{Nonparametric Estimation of the Random Coefficients Model in Python}
\author{\begin{tabular}{ccc}
Emil Mendoza\footnote{School of Mathematics and Statistics, University of Canterbury, Private Bag 4800, Christchurch 8140, New Zealand, emil.mendoza@pg.canterbury.ac.nz} \footnote{Corresponding author} &
Fabian Dunker\footnote{School of Mathematics and Statistics, University of Canterbury, Private Bag 4800, Christchurch 8140, New Zealand, fabian.dunker@canterbury.ac.nz} & Marco Reale\footnote{School of Mathematics and Statistics, University of Canterbury, Private Bag 4800, Christchurch 8140, New Zealand, marco.reale@canterbury.ac.nz}\\
\small{University of Canterbury}  & \small{University of Canterbury}  & \small{University of Canterbury}
\end{tabular}
}
\maketitle

\begin{abstract}
	\noindent 
	We present \textbf{PyRMLE}, a Python module that implements Regularized Maximum Likelihood Estimation for the analysis of Random Coefficient models. \textbf{PyRMLE} is simple to use and readily works with data formats that are typical to Random Coefficient problems. The module makes use of Python's scientific libraries \textbf{NumPy} and \textbf{SciPy} for computational efficiency. The main implementation of the algorithm is executed purely in Python code which takes advantage of Python's high-level features. 
\end{abstract}

\section{Introduction} \label{sec:method}

The Random Coefficients model is often used to model unobserved heterogeneity in a population, an important problem in econometrics and statistics. The model is given by

\begin{equation}\label{eqn:rc_model}
	Y_i=\beta_{0i}+\beta_{1i} X_{1i}+\beta_{2i} X_{2i}+\ldots+\beta_{di} X_{di}.
\end{equation}

Where $\bX_i = (1, X_{1i}, X_{2i}, \ldots X_{di})^\top$ and  $\bbeta_i = (\beta_{0i}, \beta_{1i}, \beta_{2i}, \ldots, \beta_{di})^\top$ are the regressors, and regression coefficients respectively. It is assumed that $\bX_i$, $Y_i$, $\bbeta_i$ are i.i.d random variables with $\bbeta_i$ and $\bX_i$ being independent, and that $Y_i$ and $\bX_i$ are observed while $\bbeta_i$ is unknown. 

In this paper we introduce an open source \texttt{Python} module named \pythoninline{PyRMLE} which implemenents regularized maximum likelihood estimation (RMLE) to nonparametrically estimate the density $f_\bbeta$. 

Nonparametric estimation of the random coefficients model was first established by \cite{Beran1992} for a single regressor. A kernel method for estimation was developed by \cite{Beran1996}. The optimal rate for design densities with Cauchy-type tails is derived in \cite{hoderlein2010} for a kernel method. 

Equality constrained linear regression methods were developed by \cite{fox2011simple, heiss2021nonparametric}. Regularized maximum likelihood methods have been considered for problems other than the random coefficients model specified in \eqref{eqn:rc_model} in \cite{WH:12,HW:13, hohage2016inverse, Dunker2014}. The method implemented by the \pythoninline{PyRMLE} module is the one developed in \cite{dunker2021regularized}.

\texttt{Python} was chosen as the programming language mainly for the extensive scientific and computational libraries at its disposal, namely: \pythoninline{NumPy, SciPy} by \cite{harris2020array,jones2001scipy}. The module takes advantage of the benefits of working with \pythoninline{NumPy} arrays in terms of computational efficiency achieved by doing array-wise computation. Another advantage is that there are no software imposed limits in terms of array size. The maximum array size in \texttt{Python} is solely determined by the amount of RAM available to the user, which allows the user the flexibility to increase the computational complexity of the method to the level that their system allows. The module also uses a trust-region constrained minimization algorithm developed by \cite{byrd1999interior} which is implemented in \pythoninline{SciPy}.

The paper is organized as follows: Section \ref{sec:method} briefly describes the regularized maximum likelihood method developed in \cite{dunker2021regularized}, Section \ref{sec:PythonImplementation} discusses the classes and functions available to the module, and Section \ref{sec:example} discusses examples of the modules usage for general cases.

\section{Regularized Maximum Likelihood} \label{sec:method}
 
It is assumed that the random coefficients, $\bbeta$, have a Lebesgue density $f_\bbeta$. If the conditional density $f_{Y|\bX}$ exists, the two densities are connected by the integral equation

\begin{align*}
	f_{Y|\bX}(y|\bX=\bx)=\int_{\R^{d}}\mathbbm{1}\big\{\bb^\top \bx=y\big\}f_\bbeta(\bb)d\mu_d(\bb) = \int_{\bb^\top \bx=y}f_\bbeta(\bb)d\mu_d(\bb)
	\label{eqn:line_int_f_beta}
\end{align*}

This connection allows us to employ maximum likelihood estimation to nonparametrically identify $f_\bbeta$ as seen in the following expression of the log-likelihood

\begin{equation*}
	\bar{\ell}(f_\bbeta|Y,\bX)=\frac{1}{n}\sum_{i=1}^{n}\log\left[\int_{\mathbbm{R}}\mathbbm{1}\big\{\bbeta_i^\top \bX_i=Y_i\big\}f_\bbeta(b)d\mu(\bb)\right].
	\label{eqn:avg_lhood}
\end{equation*}

Direct maximization of $\bar\ell(f_\bbeta|Y,\bX)$ over all densities it not feasible due to ill-posedness and will lead to overfitting. We stabilize the problem by adding a penalty term $\alpha \calR(f_\bbeta)$ and state the estimator as a minimization problem with negative log-likelihood
\begin{align}\label{eqn:method}
	\hat {f_\bbeta}_\alpha = \argmin_{f\ge 0,\, \|f\|_{L^1}=1} -\bar\ell(f|Y,\bX)+\alpha \calR(f).
\end{align}

Here $\alpha \ge 0$ is a regularization parameter that controls a bias variance trade-off. It was poined out in \cite{heiss2021nonparametric} that the constraint $\|f_\bbeta\|_{L^1}=1$ together with a finite difference discretization of $f_\bbeta$ is equivalent to an $\ell^1$ penalty on the discretized values of $f_\bbeta$, which could cause unwanted shrinkage of the estimate. To reduce this LASSO effect, \cite{heiss2021nonparametric} introduced an additional quadratic constraint which turns the method into an elastic net. In \cite{dunker2021regularized} it was stated that the regularization term $\alpha\calR(f_\bbeta)$ is analogous to this additional constraint but is more flexible as the method is not limited to quadratic $\calR$.

The implemented regularization terms $\calR$ in this module are: (1) squared $L^2$ norm $\calR(f) = \|f\|_{L^2}^2 = \|f\|_{2}^2$, (2) the Sobolev Norm for $H^{1}$ $ \calR(f) =  \|f_\beta\|_{2}^{2}+\|f^{\prime}_\beta\|_{2}^{2}$, and (3) entropy $\calR(f) = \int{f(\bb)\ln{f(\bb)}d\bb}$.

In addition to the regularization functional, a regularization parameter $\alpha$ also needs to be chosen. In this module we implement two methods of estimating $\alpha$: Lepskii's Balancing principle, and K-fold cross validation.

\section{Python Implementation} \label{sec:PythonImplementation}

The \pythoninline{PyRMLE} module's implementation of regularized maximum likelihood is limited to applications with up to two regressors for the random coefficients model with intercept, and up to three regressors for a model without intercept.

There are two main functions used to implement regularized maximum likelihood estimation using \pythoninline{PyRMLE}, namely: \pythoninline{transmatrix()} and \pythoninline{rmle()}. There are other sub-functions necessary to the implementation, these will be discussed under the appropriate subsections when relevant. 

\subsection{The \texttt{transmatrix()} Function}\label{sec:trans_matrix}

The purpose of the function \texttt{transmatrix()} is to construct the discrete version of the linear operator given by

\begin{equation}
	 T f_\beta = \int_{\bb^\top \bx=y}f_\bbeta(\bb)d\mu_d(\bb)
	\label{eq:inv_prob}.
\end{equation}

The linear operator above describes the integral of $f_\bbeta$ over the hyperplanes parametrized by the sample points $\bX, \text{ and }Y$. The function makes use of a finite-volume method as a discrete analog in evaluating the integral.   The function is used as follows
\begin{lstlisting}
trans_matrix = transmatrix(sample,grid)
\end{lstlisting}

The argument \texttt{sample} corresponds to the sample observations. The sample data should be in the following format: \\

\begin{center}
$\begin{bmatrix}
	X_{0,1} & X_{1,1} & \hdots & Y_1 \\
	X_{0,2} & X_{1,2} & \hdots & Y_2 \\
	\vdots & \vdots & \ddots & \vdots \\
	X_{0,n} & X_{1,n} & \hdots & Y_n \\
\end{bmatrix}$
\end{center}

In the case of a random coefficients model with intercept the first column would simply be $\bX_0 = (1,1,\hdots,1)^{T}$.

The \texttt{grid} argument is a class object generated by the \texttt{grid\_set()} function. It has the following attributes and methods: \{ \texttt{scale}, \texttt{shifts}, \texttt{interval},  \texttt{dim}, \texttt{step}, \texttt{start}, \texttt{end}, \texttt{ks()}, \texttt{numgridpoints()}\}. The \texttt{grid\_set()} function is used as follows:

\begin{lstlisting}
grid_beta = grid_set(num_grid_points=20,dim = 2)
\end{lstlisting}

The base grid that is generated by the \texttt{grid\_set()} function is a symmetric grid that spans $[-5,5]$ in each axis. The user inputs the number of grid points by passing an integer value to the function as \pythoninline{num_grid_points}. This specifies the step size of the grid as $\frac{10}{k}$ where $k$ is the number of grid points along each axis. Additionally, the user can change the range over which each axis is defined by supplying new axes ranges through the arguments: {\pythoninline{B0_range, B1_range, B2_range}} which are passed as lists or arrays that contain the end points of the new range (e.g. \pythoninline{B0_range = [0,10]} ). This is especially useful if the user expects a random coefficient to be significantly larger or smaller than the other random coefficients.

\begin{python}
grid_beta_shifted = grid_set(num_grid_points = 20, \ 
dim = 2, B0_range = [0,10])
print(grid_beta_shifted.shifts)
[-5,0,0]
\end{python}

The output of the \texttt{transmatrix()} function is the \texttt{`tmatrix'} class object that has the following attributes and methods: \\

\noindent \texttt{Tmat}: returns a $n \times m $ \pythoninline{NumPy}-array that is produced by the function \texttt{transmatrix\_2d()} or \texttt{transmatrix\_3d()}. \\
\noindent \texttt{grid}: returns the \pythoninline{class grid_obj}. \\
\noindent \texttt{scaled\_sample}: returns the scaled and shifted sample.\\
\noindent \texttt{sample}: returns the original sample. \\
\noindent \texttt{n()}: returns the number of sample observations. \\
\noindent \text{m()}: returns the number of grid points $\hat{f}_\bbeta$ is to be estimated over. 

\subsubsection{\texttt{transmatrix\_2d()}}
The 2-dimensional implementation of this method works for the random coefficients model with a single regressor and random intercept

\begin{equation}
	y_i={\beta_0}_i+{\beta_1}_{i}{x_1}_i,
	\label{eqn:rc_model_2d}
\end{equation}

and the model with two regressors and no intercept.

\begin{equation}
	y_i={\beta_1}_{i}{x_1}_i+{\beta_2}_{i}{x_2}_i+\epsilon_i
	\label{eqn:rc_model_2d_no_intercept}
\end{equation}

The function first initializes an $n \times m$-dimensional array of zeros, where $n$ is the sample size, and $m$ is the  number of grid points $f_\bbeta$ is to be estimated over. In the 1-dimensional case, the hyperplanes which $f_\bbeta$ is integrated over reduce to lines which simplifies the problem. A finite-volume type method of estimation is employed to approximate the integral expression as seen in \ref{eq:inv_prob}. This method is akin to the algebraic reconstruction methods used in Computed Tomography. Specifically, it is reminiscent to the discrete Radon Transform methodology laid out by \cite{beylkin1987discrete}.

To implement this finite-volume method, the lines parametrized by the sample points given by equations \eqref{eqn:rc_model_2d} or \eqref{eqn:rc_model_2d_no_intercept} are made to intersect with the grid. The length of each line segment that intersects with the grid is then calculated and stored in an object. The intersection points of these lines with the grid are retrieved and subsequently mapped to their respective array indices. These indices are used to map the length of the line segments to the appropriate entry in the initialized array forming the discretized linear operator $T$. The algorithm is outlined as follows:\\

\begin{algorithm}[H]
	\SetAlgoLined
	\SetKw{Input}{Input:}
	\SetKw{Initialization}{Initialization}
	\Input{sample, grid} \\
	\Initialization{Initialize \textbf{NumPy} Array of Zeros}\\
	\For{s in sample}{
		1. get intersection points\;
		2. get line segment lengths\;
		3. map intersection points to their array indices\;
		\For{i in indices}{\
			map line segment lengths to initialized array using index i
		}
	}
	\caption{\texttt{transmatrix\_2d()}}
\end{algorithm}

\medskip

\begin{figure}[!htb]
\begin{tikzpicture}[scale=0.8,every node/.style={minimum size=0.5cm},on grid]
	
	\begin{scope}[
		yshift=-50,every node/.append style={
			yslant=0.5,xslant=-1.3},yslant=0.5,xslant=-1.3
		]
		\fill[white,fill opacity=0.9] (-3,-3) rectangle (3,3);
		\draw[step=0.5cm, thin, gray] (-3,-3) grid (3,3); 
		\draw[black,very thick] (-3,-3) rectangle (3,3);
		\draw[red,ultra thick,solid] (-1.5,-3.5) -- (2.25,3.5);
		\pgfkeys{/pgf/number format/.cd, fixed, zerofill, precision =1} 
		\coordinate (s1) at (0.9107,1);
		\node at (s1) [fill=black,circle,scale=0.1] {$a$};
		\coordinate (s2) at (0.64285714285,0.5);
		\node at (s2) [fill=black,circle,scale=0.1] {$b$};
		\coordinate (s4) at (2.75,2.75);
		\node at (s4) [fill=black,circle,scale=0.15] {$c$};

	\end{scope},
	\coordinate (p1) at (-2.75,1);
	\node at (p1) [fill=black,circle,scale=0]{};
	\coordinate (p2) at (-2.75,1.5);
	\node at (p2) [fill=black,circle,scale=0]{};
	\draw[-latex,thick](-2,1)node[left,scale=1]{$P_1$}
	to[out=0,in=90] (s1);
	\draw[-latex,thick](-2,1.5)node[left,scale=1]{$P_2$}
	to[out=0,in=90] (s2);
	\draw[-latex,thick](-6,1.5)node[left,scale=1]{}
	to[out=0,in=180] (p1);
	\draw[-latex,thick](-6,1.5)node[left,scale=1]{$l_{i,j}=\|P_1 P_2\|$}
	to[out=0,in=180] (p2);
	\coordinate (s3) at (-2.75,1);
	\node at (s3) [fill=black,circle,scale=0]{};
	\node  {} {node (T) at (-4.5,-8) {%
			$\begin{aligned}
				\mathbf{T}= 
			\end{aligned}$}};
	\node  {} {node (line) at (5,-3.75) {%
		$\begin{aligned}
			\beta_0 = y_i - \beta_1 x_{1_i}
		\end{aligned}$}};
	
    \matrix (m1) at (0,-8) [matrix of math nodes,left delimiter=(,right delimiter=)](A) { 
	l_{1,1} & l_{1,2} & \dots  & l_{1,j} & \dots & l_{1,m}\\
	l_{2,1} & l_{2,2} & \dots  & l_{2,j} & \dots & l_{2,m}\\  
	\vdots  & \vdots  &  & \vdots  &  & \vdots\\
	l_{i,1} & l_{i,2} & \dots  & l_{i,j} & \dots & 0\\
	\vdots  & \vdots  &  & \vdots  &  & \vdots\\
	l_{n,1} & l_{n,2} & \dots  & l_{n,j} & \dots & l_{n,m}\\
	};
	\coordinate (lij) at (0.25,-8);
	\node at (lij) [fill=black,circle,scale=0]{};
	\draw[-latex,dashed](-8.8,1.1)node[left,scale=1]{}
	to[out=270,in=90] (lij);
	\coordinate (p3) at (2.75,2.75);
	\node at (p3) [fill=black,circle,scale=0]{};
	\draw[-latex,thick](5,2.75)node[right,scale=1]{$l_{i,m} = 0$}
	to[out=200,in=60] (s4);
	\coordinate (lim) at (3,-8.4);
	\node at (lim) [fill=black,circle,scale=0]{};
	\draw[-latex,dashed](7,2.8)node[left,scale=1]{}
	to[out=0,in=0] (lim);
	
\end{tikzpicture}
\caption{2-D Transformation Matrix Algorithm} \label{fig:1d_Algo}
\end{figure}
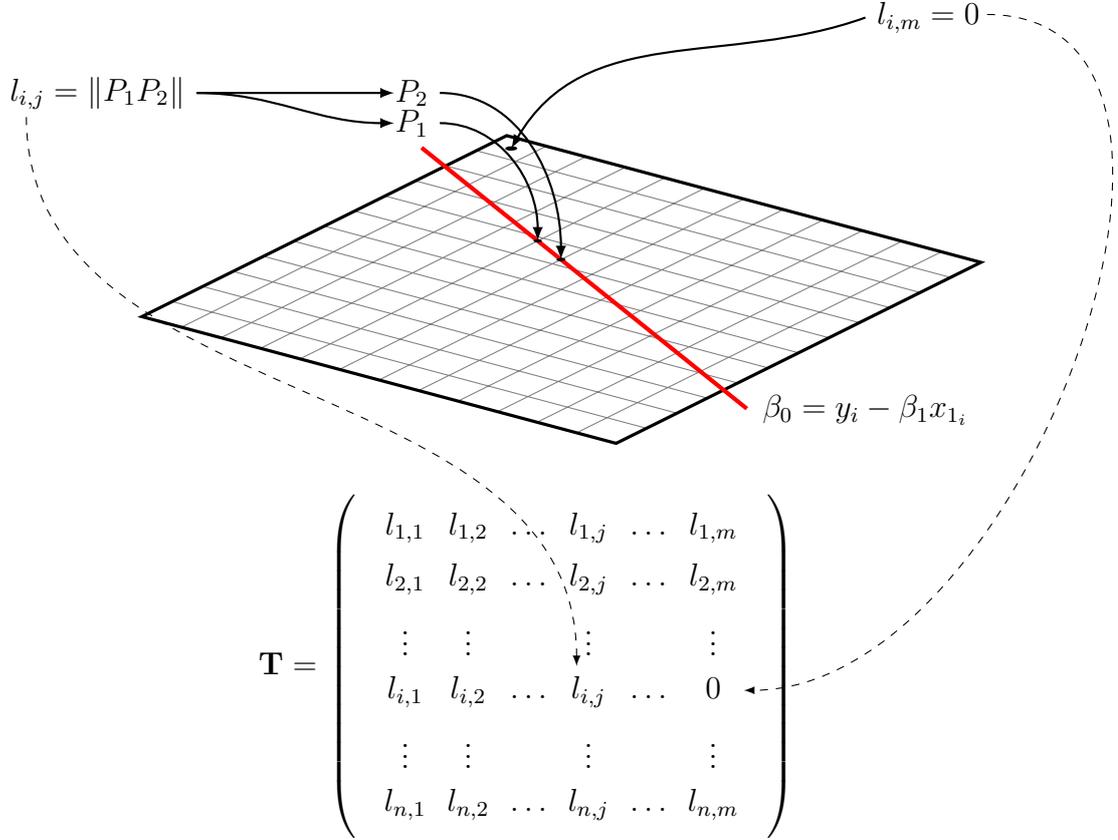

The result of this function is a large, sparse array, $\mathbf{T}$ where each row corresponds to a collection of all the lengths of the line segments intersecting the grid for a line parametrized by a sample point, i.e. each $l_{ij}$ corresponds to the length of the line segment intersecting the grid at section $i,j$. The resulting array is sparse because $l_{ij}$ is equal to zero unless a line passes through a section of the grid. The algorithm's implementation is illustrated in figure \ref{fig:1d_Algo}. 

The resulting array is then used to evaluate the log-likelihood functional to be optimized

\begin{equation}
\bar{\ell}(f_\bbeta|Y,\bX)= \frac{1}{n}\sum_{i=1}^{n}\log\textbf{T} f_\bbeta^{*},
\label{eqn:discrete_loglikelihood}
\end{equation}

where $f_\bbeta^{*} = (f_{\bbeta_{1}}, f_{\bbeta_{2}}, \hdots, f_{\bbeta_{m}})^T$ is a $m \times 1$-array or vector that serves as the discrete approximation of $f_\bbeta$.

\subsubsection{\texttt{transmatrix\_3d()}}

The implementation of the function for the 3-dimensional case works for problems with two regressors and a random intercept,

\begin{equation}
	y_i={\beta_0}_i+{\beta_1}_{i}{x_1}_i+{\beta_2}_{i}{x_2}_i,
	\label{eqn:rc_model_3d}
\end{equation}

and the model with three regressors and no intercept.

\begin{equation}
	y_i={\beta_1}_{i}{x_1}_i+{\beta_2}_{i}{x_2}_i+{\beta_3}_{i}{x_3}_i + \epsilon_i
	\label{eqn:rc_model_3d_no_intercept}
\end{equation}

Note: for the three-dimensional implementation of the algorithm numerical instabilities occur when the underlying joint density, $f_\bbeta$, has a single mode located at the center of the grid. This problem can occur in two ways: (1) it can occur organically if the mode of the density is located close to or at $(0,0,0)$ with the grid axes ranges at their default values [-5,5], or (2) artificially if the user provides grid ranges such that the mode of $f_\bbeta$ is located at its center.
This problem is overcome by simply imposing a shift onto the resulting density by applying a linear transformation to the sample data, as below:

\begin{equation}
Y_i = (\beta_{0_i} + c) + \beta_{1_i} X_1 + \beta_{2_i} X_2.
\label{eqn:sample_shift}
\end{equation}

This can be achieved in two ways: (1) the more computationally efficient method is to simply supply a grid range for $\beta_0$ that offsets the mode of $\hat{f_\bbeta}$ from the center, or (2) apply the shifting algorithm described in \eqref{eqn:sample_shift} which allows the user to supply any grid range but with an additional computational cost.

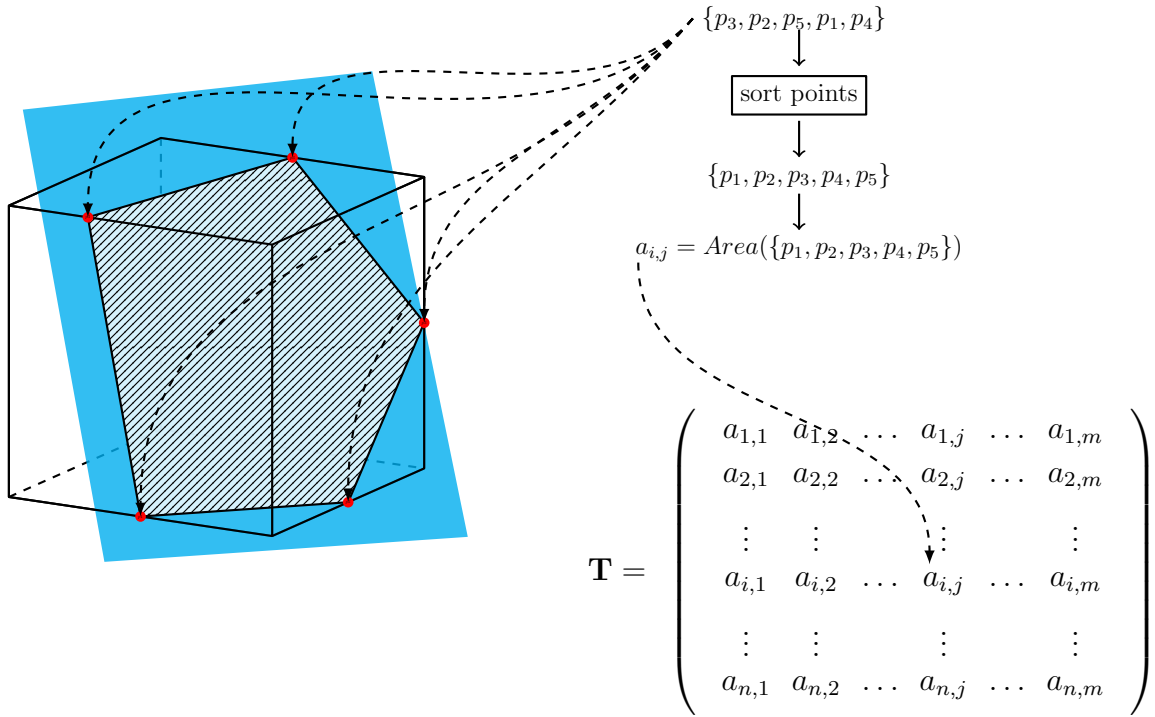
\begin{figure}[!htb]
	\begin{tikzpicture}[3d view={120}{15},line join=round, thick,
	declare function={a=4;b=2;}]
	\draw[dashed] (a,0,-a) -- (0,0,-a)-- (0,a,-a);
	\draw[dashed] (0,0,0) -- (0,0,-a); 
	\draw (a,0,0) -- (a,0,-a) -- (a,a,-a) -- (a,a,-b);
	\draw (a,a-b,0) -- (a,0,0) -- (0,0,0) -- (0,a,0) -- (a-b,a,0);
	\draw (0,a,0) -- (0,a,-a) -- (a,a,-a);
	\draw (a-b,a,0) -- (a,a,0) -- (a,a-b,0);
	\draw (a,a,0) -- (a,a,-b); 
	\fill[cyan, opacity=0.8] (4.5,0.5,1.5) -- (0.5,3.5,1.5) -- (2.75,6.25,-4) -- (6.25,2.75,-4) -- cycle;
	\draw[pattern=north east lines, preaction={fill=white, opacity=0.8}] (a,a-0.5*a,-a) -- (a-0.5*a,a,-a) -- (0,a,-a+0.5*a) -- (0,a-0.5*a,0) -- (a,a-0.7*a,0) -- cycle;
	\draw (a,0,0) -- (a,0,-a) -- (a,a,-a) -- (a,a,-b);
	\draw (a,a-b,0) -- (a,0,0) -- (0,0,0) -- (0,a,0) -- (a-b,a,0);
	\draw (0,a,0) -- (0,a,-a) -- (a,a,-a);
	\draw (a-b,a,0) -- (a,a,0) -- (a,a-b,0);
	\draw (a,a,0) -- (a,a,-b);
	\coordinate (p1) at (a,a-0.5*a,-a);
	\node at (p1) [fill=red,circle,scale=0.15] {$p1$};
	\coordinate (p2) at (a-0.5*a,a,-a);
	\node at (p2) [fill=red,circle,scale=0.15] {$p2$};
	\coordinate (p3) at (0,a,-a+0.5*a);
	\node at (p3) [fill=red,circle,scale=0.15] {$p3$};
	\coordinate (p4) at (0,a-0.5*a,0);
	\node at (p4) [fill=red,circle,scale=0.15] {$p3$};
	\coordinate (p5) at (a,a-0.7*a,0);
	\node at (p5) [fill=red,circle,scale=0.15] {$p3$};
	\draw[-latex,dashed](-8.8,3)node[right,scale=1]{}
	to[out=230,in=90] (p1);
	\draw[-latex,dashed](-8.8,3)node[right,scale=1]{}
	to[out=230,in=90] (p2);
	\draw[-latex,dashed](-8.8,3)node[right,scale=1]{}
	to[out=230,in=90] (p3);
	\draw[-latex,dashed](-8.8,3)node[right,scale=1]{}
	to[out=230,in=90] (p4);
	\draw[-latex,dashed](-8.8,3)node[right,scale=0.8]{$\{p_3,p_2,p_5,p_1,p_4\}$}
	to[out=230,in=90] (p5);
	\draw[->] (-9,4.5,0) --  (-9,4.5,-0.5);
	\coordinate (sort1) at (-9,4.5, -0.9);
	\node[draw, scale=0.8] at (sort1) {sort points};
	\draw[->] (-9,4.5,-1.25) --  (-9,4.5,-1.75);
	\node {} {node[scale=0.8] (T) at (-9,4.5,-2) {%
			$\begin{aligned}
				\{p_1,p_2,p_3,p_4,p_5\}
			\end{aligned}$}};
	\draw[->] (-9,4.5,-2.25) --  (-9,4.5,-2.75);
	\node {} {node[scale=0.8] (T) at (-9,4.5,-3) {%
			$\begin{aligned}
				a_{i,j} = Area(\{p_1,p_2,p_3,p_4,p_5\})
			\end{aligned}$}};
	\matrix (m1) at (-12,4.5,-8) [matrix of math nodes,left delimiter=(,right delimiter=)](A) { 
		a_{1,1} & a_{1,2} & \dots  & a_{1,j} & \dots & a_{1,m}\\
		a_{2,1} & a_{2,2} & \dots  & a_{2,j} & \dots & a_{2,m}\\  
		\vdots  & \vdots  &  & \vdots  &  & \vdots\\
		a_{i,1} & a_{i,2} & \dots  & a_{i,j} & \dots & a_{i,m}\\
		\vdots  & \vdots  &  & \vdots  &  & \vdots\\
		a_{n,1} & a_{n,2} & \dots  & a_{n,j} & \dots & a_{n,m}\\
	};
	\node  {} {node (T) at (-12,0,-8.75) {%
			$\begin{aligned}
				\mathbf{T}= 
			\end{aligned}$}};
	\coordinate (aij) at (-12,4.75,-8);
	\draw[-latex,dashed](-6.5,3.5,-2.75)node[right,scale=1]{}
	to[out=270,in=90] (aij);
\end{tikzpicture}
\caption{3-D Transformation Matrix Algorithm} \label{fig:2d_Algo}
\end{figure}

Similar to the 2-dimensional implementation, a finite volume type approach is used to create the discrete version of the linear operator $T$. In this higher-dimensional case the hyper-planes parametrized by the sample observations are now 2-dimensional planes, and the grid that $f_\bbeta$ is estimated over is comprised of three axes and is therefore in the shape of a 3-dimensional cube. In this case, the intersection of the plane with the grid are characterized by a collection of points that define a polygon whose areas are used as the 2-dimensional analog of the line segments in the lower dimensional implementation of this finite volume estimation process. The algorithm is outlined below. Figure \ref{fig:2d_Algo} also illustrates the algorithm for a single cuboidal grid section.\\

\begin{algorithm}[H]
	\SetAlgoLined
	\SetKw{Input}{Input:}
	\SetKw{Initialization}{Initialization}
	\Input{sample, grid} \\
	\Initialization{Initialize \textbf{NumPy} Array of Zeros}\\
	\For{s in sample}{
		get intersection points\;
		\For{p in intersection points}{\
			map intersection points to array indices\;
			\For{i in indices}{\
				assign intersection point to grid location (a point can
				belong to more than one grid box)\;}}
		\For{m in assigned points}{\
			sort points to form polygon and generate the area 
		}
		\For{i in indices}{\
			map each polygon's area to the initialized array}
		}
	
	\caption{\texttt{transmatrix\_3d()}}
\end{algorithm}

The resulting array is in the same form as the one obtained using the 2-dimensional implementation \texttt{transmatrix\_2d()}, and can be used in the following mannner to evaluate the log-likelihood functional in \eqref{eqn:discrete_loglikelihood}

\begin{equation*}
	\textbf{T}f_\bbeta^{*} =
	\begin{bmatrix}
		a_{1,1} & a_{1,2} & a_{1,3} & \hdots & a_{1,m} \\
		a_{2,1} & a_{2,2} & a_{2,3} &\hdots & a_{2,m} \\
		a_{3,1} & a_{3,2} & a_{3,3} &\hdots & l_{3,m} \\
		\vdots & \vdots & \vdots & \ddots & \vdots \\
		a_{n,1} & a_{n,2} &a_{n,3} &\hdots & a_{n,m} \\
	\end{bmatrix} 
	\begin{bmatrix}
		{f_\beta}_1 \\
		{f_\beta}_2 \\
		{f_\beta}_3 \\
		\vdots      \\
		{f_\beta}_m
	\end{bmatrix},
\end{equation*}

where in this case each $a_{ij}$ corresponds to the area of the polygonal intersection of the plane with the discrete estimation grid. The resulting product of this matrix and vector is an $n \times 1$ array. Applying an element-wise log-transformation, then taking the sum of this $n \times 1$ array results in the expression in \eqref{eqn:discrete_loglikelihood}.

It is important to note the computational performance of this algorithm. Matrix multiplication parallelizes the computation and eliminates the need to evaluate the log-likelihood functional sequentially for each sample point. However, the process of creating the transformation matrix $\mathbf{T}$ scales linearly with the sample size $n$ and exponentially with the dimensionality of the problem. The function was written in a way that maximizes efficiency as much as possible by making use of Python's parallel computation through array-wise computation when possible. 

Solving for the transformation matrix, $\mathbf{T}$ is only one of two steps in the process of estimating the density $f_\bbeta$. The second step is the process of optimizing the regularized log-likelihood functional, the run-time of which scales multiplicatively with respect to the sample size, $n$, and the number of discretezation points, $m$. Therefore, choosing an appropriate sample size and level of discretezation is an important consideration as both primarily determine the total cost of running the algorithm.

\subsection{The \texttt{rmle()} Function}\label{sec:rmle_func}

The \texttt{rmle()} function returns a class named \texttt{`RMLEResult'}. This class stores the solution to the optimization problem, $f_\bbeta^{*}$, and metadata about the solution and the process of optimization. An instance of \texttt{`RMLEResult'} has the following accessible attributes and methods:\\
\noindent \texttt{f}: returns a $m \times 1$ array containing all the estimated function values. It is necessary to reshape the solution before visual representation.\\
\noindent \texttt{f\_shaped}: returns the reshaped array of $f$.\\
\noindent \texttt{dim}: returns an integer which represents how many dimensions $f_\bbeta$ is estimated over.\\
\noindent \texttt{maxval()}: returns a list containing the maximum value of $f_\bbeta^{*}$ and its location. \\
\noindent \texttt{mode()}: returns a list containing possible modes of the density and their locations. \\
\noindent \texttt{ev()}: returns atuple containing the expected value of each $\beta_i$. \\
\noindent \texttt{alpha}: returns a floating-point that specifies the regularization parameter, $\alpha$, used for estimation.\\
\noindent \texttt{alpmth}: returns a string that specifies the method by which the regularization parameter, $\alpha$ was chosen. It can take on the following values: \{`Lepskii', `CV',  `User'\}.\\
\noindent \texttt{T}: returns a class object that is created using the \texttt{transmatrix()} function. It contains attributes, methods and subclasses that contain information about the transformation matrix $\mathbf{T}$ and meta-data about it. \\
\noindent \texttt{Tmat}: returns an attribute of the sublcass \texttt{T} but also accessible from the \texttt{RMLEResult} class. It returns the transformation matrix $\mathbf{T}$. \\ 
\noindent  \texttt{grid}: returns a class object that is created using the \texttt{grid\_set()} function. It has attributes and methods that contain information about the grid $\hat{f}_\bbeta$ is estimated over. \\
\noindent \texttt{details}: returns a dictionary containing metadata about the minimization process. \\

The function \texttt{rmle()} serves as a wrapper function for \textbf{SciPy}'s \texttt{minimize} function with \texttt{`trust-const'} set as the minimization algorithm. The choice of this algorithm is crucial as it specializes in large-scale constrained minimization problems, for further details we refer to \cite{byrd1999interior}. The ability to handle large-scale problems is important because depending on the size of the sample, and level of discretezation, the functional evaluations as well as the gradient evaluations could become exceedingly expensive, as it scales with both in a multiplicative manner ($n \times m$). The option to set constraints was also an important consideration. As in equation \eqref{eqn:method} there are two important constraints in estimating $f_\bbeta$, namely: $f\ge 0,\, \text{and } \|f\|_{L^1}=1$. These two constraints ensure the resulting solution satisfies the definition of a density.

Table 1 contains all the arguments for the function \texttt{rmle()} followed by a short description of what they pertain to. More important arguments will be discussed in following subsections. \\

\begin{table}[!h]
	\caption{\texttt{rmle()} arguments}
	\centering 
	\begin{tabular}{c ll} 
		\hline\hline \\ [-1.5ex]
		\textbf{Argument} & \textbf{Description}  \\\hline
		\\ [-1.5ex]
		 \multicolumn{1}{p{3cm}}{\raggedright functional} & \multicolumn{1}{p{11cm}}{\raggedright Negative likelihood functional with corresponding regularization term} \\
		\multicolumn{1}{p{3cm}}{\raggedright alpha}  & \multicolumn{1}{p{11cm}}{\raggedright constant $ \geq 0$ that serves as the regularization parameter, or a string matching: `cv' or `lepskii'.}\\
		\multicolumn{1}{p{3cm}}{\raggedright tmat}   & \multicolumn{1}{p{11cm}}{\raggedright Class object returned by \texttt{transmatrix()} which contains information about the transformation matrix $\mathbf{T}$ and the grid it is estimated over.}  \\
		\multicolumn{1}{p{3cm}}{\raggedright k}  & \multicolumn{1}{p{11cm}}{\raggedright Optional argument: integer which specifies how many folds for modified k-fold cross-validation. Default value is $k = 10$ }\\
		\multicolumn{1}{p{3cm}}{\raggedright initial\_guess}  & \multicolumn{1}{p{11cm}}{\raggedright Optional argument from the \textbf{SciPy Optimize} minimize function. Used to supply an initial value for the minimization algorithm to start. Default value is set to a \textbf{NumPy} array of values close to zero.} \\
		\multicolumn{1}{p{3cm}}{\raggedright hessian\_method} & \multicolumn{1}{p{11cm}}{\raggedright Optional argument from \textbf{SciPy Optimize} minimize function. Default value is set to `2-point'.} \\
		\multicolumn{1}{p{3cm}}{\raggedright constraints} & \multicolumn{1}{p{11cm}}{\raggedright Refers to the linear constraints imposed onto the problem. It is set as an optional argument, the default value is set to $\|f\|_{L^1}=1$} \\
		\multicolumn{1}{p{3cm}}{\raggedright tolerance}  &  \multicolumn{1}{p{11cm}}{\raggedright Optional argument for the tolerance criteria for the optimization algorithm's termination. Default value is set to \texttt{1e-6}.} \\
		\multicolumn{1}{p{3cm}}{\raggedright max\_iter}  & \multicolumn{1}{p{11cm}}{\raggedright Optional argument for the maximum number of iterations. Default value is set to 100.} \\
		\multicolumn{1}{p{3cm}}{\raggedright bounds}  & \multicolumn{1}{p{11cm}}{\raggedright Refers to the bound constraints of the optimization problem. The default is expressed as $f\ge 0$ } \\
		\hline
	\end{tabular}
	\label{tab:sobolev_args}
\end{table}
\subsubsection{Functionals and Regularization Terms}

Recall the average log-likelihoood functional to be minimized as in equation \eqref{eqn:method}.  As specified in section \ref{sec:method} the regularization terms implemented in this module are: the Sobolev norm for $H^1$, the squared $L^2$ norm, and Entropy. The module also includes an option for a functional that has no regularization term if the user wishes to produce a reconstruction of $f_\bbeta$ without any form of regularization. This option will often lead to overfitting, and produce a highly unstable solution.

\subsubsection{Sobolev Norm for $H^1$}

The functional incorporating the Sobolev norm for $H^1$ has the following form,
\begin{align} \label{eqn:h1_penalty}
	-\bar{\ell}(f_\bbeta|Y,\bX)+\alpha (\|f\|_{2}^{2}+\|f^{\prime}\|_{2}^{2})
\end{align}

where $\|\cdot\|_{2}^{2}$ indicates the squared $L^2$ norm. It is important to note that the choice of the regularization term would typically depend on knowledge about the true solution, as the choice of the regularization term imposes certain assumptions on $f_\bbeta$. In the case of the $H^1$ penalty term, the solution has a square-integrable first derivative in addition to the assumptions of non-negativity and integrability imposed by the constraints of the minimization problem. 

The function \pythoninline{sobolev()}, or \pythoninline{sobolev_3d} for the 3-dimensional application returns the value of the discrete implementation of the functional in \eqref{eqn:h1_penalty}. Table 2 provides details on the arguments required for this function. The \textbf{SciPy Optimize} minimize function does not accept array arguments that are greater than one dimension. A necessary step is to unravel the transformation matrix, $\mathbf{T}$, into a one-dimensional array which is passed to the function as \texttt{tm\_long} and simply reshaped into the proper array dimensions. The term $\mathbf{T}f$ is calculated using \textbf{NumPy's} matrix multiplication and sum functions which are much faster alternatives than their non-\textbf{Numpy} counterparts. The regularization term is calculated  in \eqref{eqn:h1_penalty}, with the function \texttt{norm\_fprime()} computes for $\|f^{\prime}\|_{2}^{2}$ where $f^{\prime}$ is treated as a total derivative.

\begin{table}[!h]
	\caption{\texttt{sobolev(), sobolev\_3d()} arguments}
	\centering 
	\begin{tabular}{c rrrr} 
		\hline\hline \\ [-1.5ex]
		Argument & Notation & Description \\ [0.5ex]
		\hline 
		f & $f_\bbeta^{*}$ & current value of the solution $f_\beta^{*}$ & \\ 
		a & $\alpha$ & constant that serves as the regularization parameter & \\
		tm\_long & $\mathbf{T}$ & unraveled form of the transformation matrix, $\mathbf{T}$ &\\
		n & n & the sample size & \\
		s & $\Delta b$ & the step size of the grid & \\
		\hline 
	\end{tabular}
	\label{tab:sobolev_args}
\end{table}

The underlying minimization function is able to approximate the Jacobian of the functional with an additional computational cost. For computational efficiency, we supply an explicit form for the Jacobian of the functional: $-\frac{\bar{\ell}(f_\bbeta|Y,\bX)^{\prime}}{\bar{\ell}(f_\bbeta|Y,\bX)}+2\alpha (f-f^{\prime\prime})$

The form of this Jacobian implies an additional smoothness assumption on the solution, as it requires $f_\beta$ to be twice differentiable. 

\subsubsection{Squared $L^2$ Norm}

The form of the functional in \eqref{eqn:method} that incorporates the squared $L^2$ norm as the regularization term is:
\begin{align*}
	-\bar{\ell}(f_\bbeta|Y,\bX) f+\alpha\|f\|_{2}^{2}
\end{align*}
The arguments for this function are indentical to those listed in Table \ref{tab:sobolev_args}, likewise is true for the Jacobian associated with this regularization term. The Jacobian has the following form: $-\frac{\bar{\ell}(f_\bbeta|Y,\bX)^{\prime}}{\bar{\ell}(f_\bbeta|Y,\bX)}+2\alpha f$

Choosing this regularization functional imposes less smoothness assumptions on the solution, as the only additional assumption in place is square-integrability. This leads to a typically less smooth reconstruction as compared to using the Sobolev norm for $H^1$ as the regularization functional. The functions in python are coded similarly as with the $H^1$ regularization functional.

\subsubsection{Entropy}
The form of the functional in \eqref{eqn:method} that incorporates the entropy of the function has the following form: $-\bar{\ell}(f_\bbeta|Y,\bX)+\alpha \int f \log(f)db$
This functional has the least amount of assumptions on the solution, $f_\bbeta$. It only requires finite entropy which is a weak assumption in addition to the non-negativity and $L^1$ constraints of the minimization problem.

The Jacobian of the entropy functional also does not impose any additional assumptions on the solution, and has the following form: $-\frac{\bar{\ell}(f_\bbeta|Y,\bX)^{\prime}}{\bar{\ell}(f_\bbeta|Y,\bX)}+\alpha (\log f+1)$

\subsubsection{Parameter Selection}
Recall the minimization problem as in \eqref{eqn:method} where a constant $\alpha \geq 0$ controls the size of the effect of the regularization term. The user can provide the $\alpha$ value directly if the user has a general idea of the level of smoothing necessary for the solution. If the user has no best-guess for the value of the regularization parameter alpha, the module has two options to automatically select the parameter $\alpha$, namely: Lepskii's balancing principle, and k-fold cross-validation. The Lepskii method typically yields a less accurate result relative to k-fold cross-validation; however, its advantage lies in significantly less computational cost.

\subsubsection{Lepskii's Balancing Principle}

Lepskii's principle is an adaptive form of estimation which is popular for inverse problems, e.g. \cite{Tsybakov:00}, \cite{BauHoh:05}, \cite{mathe_2006}, \cite{hohage2016inverse}, \cite{Werner:18}. It is significantly computationally less expensive than other parameter selection methods.

The method works as follows: we compute $\hat{f_\bbeta}_{\alpha_{1}},\ldots,\hat{f_\bbeta}_{\alpha_{m}}$ for $\alpha_1 = c_{L_{n}} \frac{\ln(n)}{\sqrt{n}}$ and $\alpha_{i+1} = r\alpha_i$ with some constants $c_{L_{n}} >0,r>1$. We then select $\alpha_{j}$ as the optimal parameter choice where:

\[
j_{bal}:=\max\{j\leq m, \|\hat{f_\bbeta}_{\alpha_i}-\hat{f_\bbeta}_{\alpha_j}\| \leq 8 r^\frac{1-i}{2}, \text{ for all } i < j\}.
\]

The algorithm is implemented in python as follows: \\

\begin{algorithm}[H]
	\SetAlgoLined
	\SetKw{Input}{Input:}
	\SetKw{Initialization}{Initialization}
	\Input{$\{\alpha_1,\alpha_2,\hdots,\alpha_m\}$} \\
	\Initialization{Generate the transformation matrix $\mathbf{T}$}\\
	\For{$\alpha_i$ in $\{\alpha_1,\alpha_2,\hdots,\alpha_m\}$}{
		Compute for $\hat{f_\bbeta}_{\alpha_{i}}$ using $\mathbf{T}$\;
	}
	\For{j in $\{1,2,\hdots,m\}$}{\
		Set $i = 0$ \;
		\While{i $<$ j}{\
			Check if $\|\hat{f_\bbeta}_{\alpha_i}-\hat{f_\bbeta}_{\alpha_j}\|\leq 8 r^\frac{1-i}{2}$ \;
			\If{$\|\hat{f_\bbeta}_{\alpha_i}-\hat{f_\bbeta}_{\alpha_j}\|> 8 r^\frac{1-i}{2}$}{\
				$j_{bal} = j$ \;
				\textbf{break}
			}
			 $i+=1$
		}
	}
	\caption{Lepskii's Balancing Principle}
\end{algorithm} 
\medskip
The bulk of the computational cost of the Lepskii algorithm implementation can be broken down into two components: the fixed cost of generating $\mathbf{T}$, and the variable cost of computing $\hat{f_\bbeta}_{\alpha_{1}},\ldots,\hat{f_\bbeta}_{\alpha_{m}}$ as the number of $\alpha$ values to be used depends on $c_{L_{n}} >0 \text{, }r>1$, and also the sample size of the data. This implementation of Lepskii algorithm's scales linearly in terms of runtime with the number of $\alpha$ values being tested, $m$.

\subsubsection{K-Fold Cross Validation}
Cross-validation is another popular parameter choice rule. Here we present a modified implementation of k-fold cross-validation with a cost-function that is applicable to our problem. The modification we apply is an algorithm to lessen the computation time by reducing the number of $\alpha$ values it needs to iterate through.
The loss function we considered was,

\begin{align*}
	J_{\alpha_{i}} = -\sum_{j=1}^{k} \log \mathbf{T}_{j}\hat{f}_{\bbeta_{-j}}
\end{align*}

Where $\mathbf{T}_{j}$ is the transformation matrix generated from a subsample of the observations, which can be interpreted as the $j$-th fold that is left out in the current iteration, and $\hat{f}_{\bbeta_{-j}}$ is the estimate for $f_\bbeta$ using $\mathbf{T}_{-j}$. The loss function can be interpreted as the negative of the likelihood that the $j$-th fold of the sample used to generate $\mathbf{T}_{-j}$ was drawn from the distribution, as the subsample fold used to produce $\mathbf{T}_{j}$. We aim to choose the $\alpha$ that minimizes this loss function.

The search method for the optimal $\alpha$ value reduces the number of $\alpha$ values tested. The algorithm involves separating the range of $\alpha$ values into two sections $\{\alpha_{1},\ldots,\alpha_j\}$ and $\{\alpha_{j+1},\ldots, \alpha_{m}\}$. Two alpha values $\alpha_a$, and $\alpha_b$ are randomly selected from the respective sections and are used to compute for the corresponding loss function values, $J_{\alpha_{a}}$ and $J_{\alpha_{b}}$. The section from which the $\alpha$ value that produces the smaller $J_{\alpha}$ was drawn from is kept, while the other is discarded. This is repeated until there is a sufficiently small range of $\alpha$ values. Once this range of $\alpha$ values is obtained, the loss function is evaluated over all the remaining $\alpha$ values and the optimal $\alpha$ is chosen as the one which minimizes $J_{\alpha}$. The complete algorithm is implemented as follows:\\

\begin{algorithm}[H]
	\SetAlgoLined
	\SetKw{Input}{Input:}
	\SetKw{Initialization}{Initialization}
	\Input{$\{\alpha_1,\alpha_2,\hdots,\alpha_m\}$} \\
	\Initialization{Generate the transformation matrix $\mathbf{T}$ and apply a random shuffle, set $\boldsymbol{\alpha}$ = $\{\alpha_1,\alpha_2,\hdots,\alpha_m\}$}\\
	\While{\texttt{len}($\boldsymbol{\alpha}$) $>$ 3}{
		1. Set $\boldsymbol{\alpha_a}$ = $\{\alpha_1,\alpha_2,\hdots,\alpha_j\}$ \;
		2. Set $\boldsymbol{\alpha_b}$ = $\{\alpha_{j+1},\alpha_{j+2},\hdots,\alpha_m\}$ \;
		3. Randomly select $\alpha_a$ and $\alpha_b$ from $\boldsymbol{\alpha_a}$ and $\boldsymbol{\alpha_b}$ respectively. \;
		4. Evaluate $J_{\alpha_{a}}$ and $J_{\alpha_{b}}$ \;
		\If{$J_{\alpha_{a}}$ $<$ $J_{\alpha_{b}}$}{	
			Set $\boldsymbol{\alpha}$ = $\boldsymbol{\alpha_a}$ 
		}
		\Else{
			Set $\boldsymbol{\alpha}$ = $\boldsymbol{\alpha_b}$	
		}
	}
	\For{$\alpha_i$ in $\boldsymbol{\alpha}$}{\
		Compute for $J_{\alpha_{i}}$ \;
	}
	Choose $\alpha_{cv} = \argmin{J_{\alpha_{i}}}$
	\caption{Modified K-fold Cross Validation}
\end{algorithm} 
\medskip

The runtime of the unmodified version of k-fold cross-validation scales linearly with the product $k \times m$ where $k$ is the number of folds and $m$ is the number of $\alpha$ values being tested. Applying the modified version reduces the number of $\alpha$ values being tested, $m$, by some logarithmic factor, which is a significant reduction in computational cost which makes cross-validation more computationally feasible.

\section{Examples} \label{sec:example}

The following examples will be demonstrated in this section: the random coefficients model with a single regressor and random intercept for the two-dimensional case, and the two regressors and random intercept for the three-dimensional case. This section will also show how to plot the estimated density using the built in plotting function \pythoninline{plot\_rmle()} which makes use of functions from the \pythoninline{matplotlib} library. It will also show how to plot the density without the use of the built-in function in case the user wishes to explore different plotting options.

\subsection{Example 1: 2-D Case Single Regressor with Random Intercept}

The first example is the case described by \eqref{eqn:rc_model_2d}. The example is demonstrated with simulated data using the function \pythoninline{sim\_sample()}. This function simulates the regressor $X_1 \sim \unif(-2,2)$ and the random coefficients $\beta_0$, $\beta_1$ from a bimodal multivariate normal mixture as follows,

\begin{align*}
	0.5\mathcal{N}([-0.5,-0.5], 0.01\mathbbm{I}_2) + 0.5\mathcal{N}([0.5,0.5], 0.01\mathbbm{I}_2).
\end{align*}

The general flow of the process of using the module can be broken down in five steps:

\begin{enumerate}[topsep=0pt,itemsep=-1ex,partopsep=1ex,parsep=1ex]
	\item Import the necessary modules and functions.
	\item Establish the dataset to be used (either real or simulated data).
	\item Specify the grid over which $\hat{f_\bbeta}$ is to be estimated over.
	\item Generate the transformation matrix $\mathbf{T}$.
	\item Run the \pythoninline{rmle()} function.
\end{enumerate}

\begin{python}
from pyrmle import *

sample = sim_sample(n = 10000,dim = 2)
\end{python}

\medskip

The program begins by importing the necessary functions from \pythoninline{pyrmle.py} which contains the high-level functions that the user interacts with. The module \pythoninline{pyrmle_funcs} contains the underlying functions necessary for functions in the main module \pythoninline{pyrmle} to run. The next step is to define the sample to be used in creating the transformation matrix $\mathbf{T}.$ The sample has the same form as described in subsection \ref{sec:trans_matrix}, where the sample has the form $[\bX_0, \bX_1, \mathbf{Y}]$. In Python it takes the shape of $10000 \times 3$ \pythoninline{NumPy} array as seen below.

\medskip

\begin{python}
	[[ 1.        , -0.76765455,  2.10802347],
 	[ 1.        ,  1.51774991, -0.11337413],
 	...,
 	[ 1.        , -1.80486996, -3.08120151]]
\end{python}

\medskip

The next step is to generate the grid over which $\hat{f_\bbeta}$ is estimated over. This is done using the \pythoninline{grid\_set()} function, as discussed briefly in \ref{sec:trans_matrix}. In this example we set the ranges of $\beta_0$ and $\beta_1$ to $[-10,10]$ which defines a two-dimensional grid spanning that range in each axis. This function creates an instance of the  \pythoninline{class grid_obj}  which has attributes and methods enumerated and described in subsection \ref{sec:trans_matrix}. When the \pythoninline{grid} class instance has been created, the user can proceed to generate an instance of the \pythoninline{class tmatrix}  using the \pythoninline{transmatrix()} function. 

\medskip

\begin{python}
grid_beta = grid_set(num_grid_points = 40, dim = 2)
print(grid_beta.numgridpoints())
1600
T = transmatrix(sample, grid_beta)
\end{python}

\medskip

After the instance of the transformation matrix is produced, the user can then run the \pythoninline{rmle()} function. As stated in subsection \ref{sec:rmle_func}, the \pythoninline{rmle()} function has three essential arguments: \{\texttt{`functional',`alpha',`tmat'}\}.
\medskip

\begin{python}
result = rmle(sobolev,0.25,T)
print(result.ev())
[-0.002865326246726808, -0.010416375635973668]
print(result.mode()[:2])
[[0.39681799850755783, [-0.625, -0.375]],
[0.38831830923870914, [0.625, 0.375]]]
plot_rmle(result)
plot_rmle(result,plt_type='surface')
	
\end{python}

\begin{figure}[h]
	\centering
	\begin{minipage}{0.45\textwidth}
		\centering
		\includegraphics[width=1.1\textwidth]{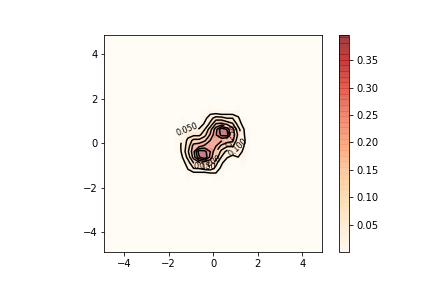} 
		\caption{$\hat{f_\beta}$ contour plot with 40 grid points}\
		\label{fig:cont_40_base}
	\end{minipage}\hfill
	\begin{minipage}{0.45\textwidth}
		\centering
		\includegraphics[width=1.1\textwidth]{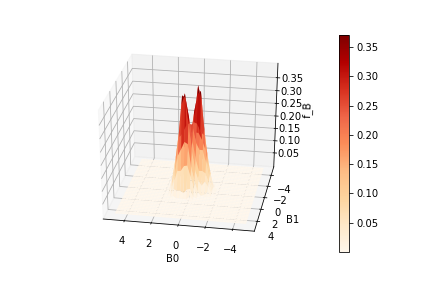} 
		\caption{$\hat{f_\beta}$ surface plot with 40 grid points}
		\label{fig:surface_40_base}
	\end{minipage}
\end{figure}

As stated in subsection \ref{sec:rmle_func}, the \pythoninline{rmle()} function generates an instance of \pythoninline{class RMLEResult}. This class has a number of useful and informative attributes and methods that describe the estimated density. The \pythoninline{ev()} method returns the expected value of each $\bbeta_j$, while the \pythoninline{mode()} method returns possible maxima of the estimated density. The \pythoninline{mode()} method relies on a naive search method for maxima with no underlying statistical tests.

Figure 3 (reference) shows the contour plot produced by the \pythoninline{plot_rmle()} function. It is clear that there is a large portion of the grid that is essentially unused, and the user could benefit from a reduction in the grid-size in terms of computational costs. This can be done by reducing the number of grid points and shrinking the range of each axis. In terms of tuning the size of the grid, we suggest that the user tries a relatively large range to begin with to ensure that the grid contains the support of $\hat{f_\bbeta}$, and then consider smaller grid ranges. Having a grid range that is too small has a negative effect on the estimate as the optimization algorithm enforces the constraint that $\|\hat{f_\bbeta}\|_{L^1}=1$.

\medskip

\begin{python}
grid_beta_alt = grid_set(20,2,B0_range=[-1.5,1.5],\
B1_range=[-1.5,1.5])
print(grid_obj_alt.numgridpoints)
400
T2 = transmatrix(sample, grid_beta_alt)
result2 = rmle(sobolev,0.15,T2)

print(result2.ev())
[-0.004977073000670898, -0.003964663139211258]
print(result2.mode()[:2])
[[0.3643228046077392, [0.5249999999999, 0.5249999999999]],
[0.3580092260598125, [-0.5249999999999, -0.5249999999999]]]
plot_rmle(result2)
plot_rmle(result2,plt_type='surface')
\end{python}

\begin{figure}[h]
	\centering
	\begin{minipage}{0.45\textwidth}
		\centering
		\includegraphics[width=1.1\textwidth]{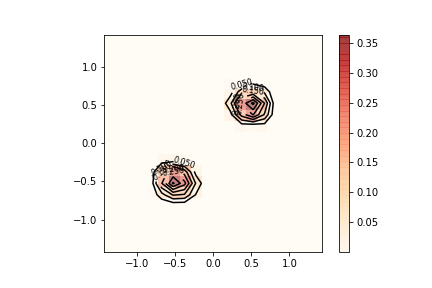} 
		\caption{$\hat{f_\beta}$ contour plot with 20 grid points on a smaller grid}\
		\label{fig:20_smaller_cont}
	\end{minipage}\hfill
	\begin{minipage}{0.45\textwidth}
		\centering
		\includegraphics[width=1.1\textwidth]{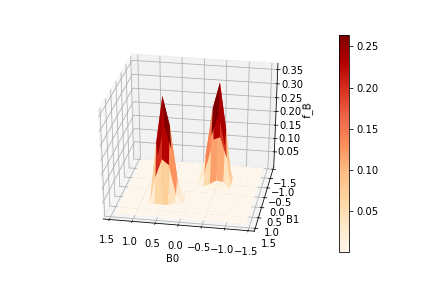} 
		\caption{$\hat{f_\beta}$ surface plot with 20 grid points on a smaller grid}
		\label{fig:20_smaller_surface}
	\end{minipage}
\end{figure}

The reduction in the number of grid points resulted in a significant reduction in the run time of the algorithm while achieving a better estimate for $\hat{f_\bbeta}$. With the reduction in the computational cost of the algorithm makes it more favorable to run an automatic parameter choice method. 

\begin{python}
result_cv = rmle(sobolev,'cv',T2)
print(result_cv.alpha)
0.07065193045869372
print(result_cv.ev())
[-0.004977073000670898, -0.003964663139211258]
print(result_cv.mode()[:2])
[[0.3643228046077392, [0.5249999999999, 0.5249999999999]],
[0.3580092260598125, [-0.5249999999999, -0.5249999999999]]]
plot_rmle(result_cv)
plot_rmle(result,plt_type='surface')
\end{python}

\begin{figure}[h]
	\centering
	\begin{minipage}{0.45\textwidth}
		\centering
		\includegraphics[width=1.1\textwidth]{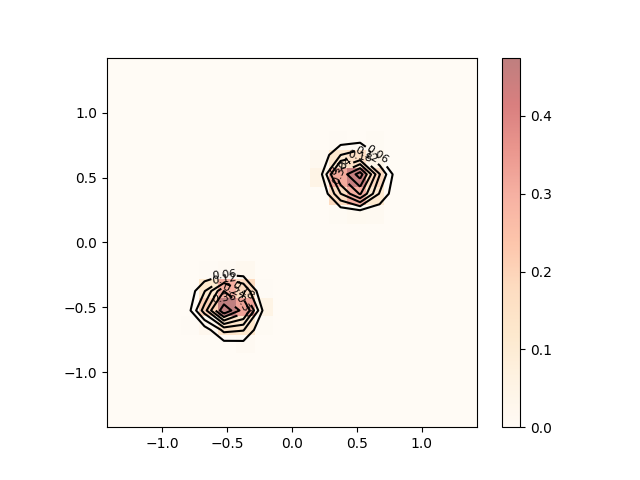} 
		\caption{$\hat{f_\beta}$ contour plot with 20 grid points on a smaller grid using cross-validation}\
		\label{fig:20_cv_cont}
	\end{minipage}\hfill
	\begin{minipage}{0.45\textwidth}
		\centering
		\includegraphics[width=1.1\textwidth]{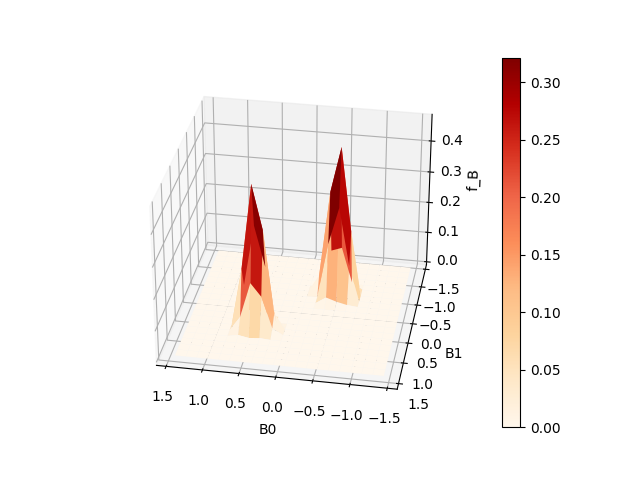} 
		\caption{$\hat{f_\beta}$ surface plot with 20 grid points on a smaller grid using cross-validation}
		\label{fig:20_cv_surface}
	\end{minipage}
\end{figure}

The general workflow that we suggest when tuning the parameters to be used in estimation is as follows:

\begin{enumerate}[topsep=0pt,itemsep=-1ex,partopsep=1ex,parsep=1ex]
	\item Establish a relatively large grid range for estimation and generate the transformation matrix $\mathbf{T}$. This should be treated as an exploratory step in terms of analyzing the data.
	\item Set $\alpha$ equal to the step size of the grid.
	\item Run the \pythoninline{rmle()} function.
	\item Plot $\hat{f_\bbeta}$ using the \pythoninline{plot_rmle()} function and determine the necessary grid range.
	\item Limit the grid range as well as the grid points to reduce computation costs and genrate the new matrix $\mathbf{T}^*$.
	\item Run the \pythoninline{rmle()} function with $\mathbf{T}^*$, and optionally employ one of the two automatic parameter selection methods: \{\pythoninline{'cv','lepskii'}\}.
\end{enumerate}

The following example will demonstrate usage of the \pythoninline{grid_set()} function in terms of supplying a different range $\beta_{1}$. The simulated data in this case will have modes for $\beta_1$ that are significantly larger than that of $\beta_0$ and are not encapsulated by the default range $[-5,5]$. The betas are sampled from the following distribution: $0.5\mathcal{N}([-1.5,6], \mathbbm{I}_2) + 0.5\mathcal{N}([1.5,9], \mathbbm{I}_2).$

\begin{python}
cov = [[[1, 0], [0, 1]],[[1, 0], [0, 1]]]
mu = [[-1.5,6],[1.5,9]]
sample = sim_sample(10000,2,beta_mu = mu,beta_cov = cov))
grid_beta_shifted = grid_set(num_grid_points = 20, \ 
dim = 2, B1_range=[2,13])
T_shifted = transmatrix(sample, grid_beta_shifted)
result_shifted = rmle(sobolev_norm_penal,0.5,T_shifted)

print(result_shifted.ev())
[0.03520704073478552, 7.524001743037029]
print(result.mode()[0:2])
[[0.07133616078580148, [1.25, 8.875]],
[0.0652140364538059, [-1.75, 5.574999999999999]]]]

plot_rmle(result_shifted)
plot_rmle(result_shifted,plt_type='surface')

\end{python}

\begin{figure}[h]
	\centering
	\begin{minipage}{0.45\textwidth}
		\centering
		\includegraphics[width=1.1\textwidth]{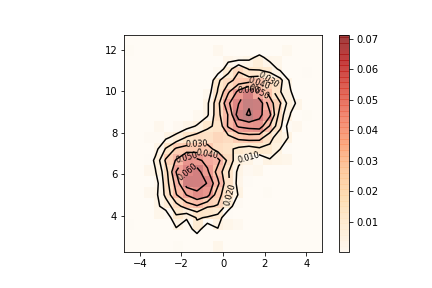} 
		\caption{$\hat{f_\beta}$ contour plot with 20 grid points on shifted grid}\
		\label{fig:20_shifted_cont}
	\end{minipage}\hfill
	\begin{minipage}{0.45\textwidth}
		\centering
		\includegraphics[width=1.1\textwidth]{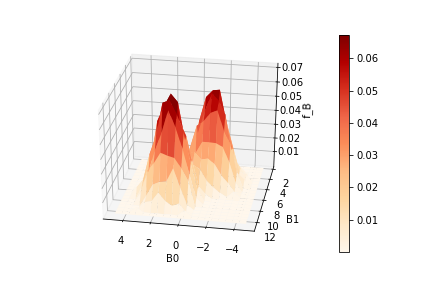} 
		\caption{$\hat{f_\beta}$ surface plot with 20 grid points on a shifted grid}
		\label{fig:20_shifted_surface}
	\end{minipage}
\end{figure}

\subsubsection{Example 2: 3-D Case Two Regressors with Random Intercept}

This second example is the case described by \eqref{eqn:rc_model_3d}. The example is demonstrated likewise with simulated data using the same function \pythoninline{sim_sample()}. The regressors are simulated as follows, $X_1, X_2$ are i.i.d $\unif(-2,2)$, and the random coefficients $\beta_0, \beta_1, \text{ and } \beta_2$ are simulated from $\mathcal{N}([2,2,2], 0.01\mathbbm{I}_3)$

\medskip

\begin{python}
from pyrmle import *
	
sample = sim_sample(n = 10000,dim = 3)
\end{python}

\medskip

As in the previous example, the program begins by importing the necessary modules and functions. The \pythoninline{sim_sample()} function generates sample observations based on the aforementioned distributions. This results in a $10,000 \times 4$ \pythoninline{NumPy} array. 

\medskip

\begin{python}
grid_beta = grid_set(10,3,B0_range=[-2,4],\ 
B1_range=[-2,4],B2_range=[-2,4])
print(grid_beta.numgridpoints())
1000
T = transmatrix(sample,grid_beta)

\end{python}

\medskip

The next step is to establish the number of grid points, and to generate the transformation using the simulated sample observations. In this case, we first consider ten grid points in each axis amounting to a total of 1000 grid points. If the user wishes to estimate $\hat{f_\bbeta}$ over a finer grid it would be more efficient to first determine the smallest possible range of each axis that would fit almost all of the probability mass of $\hat{f_\bbeta}$.

\medskip

\begin{python}
result = rmle(sobolev_norm_penal2d,0.6,T)
print(result.ev())
[2.458436489282234, 2.25500373629305, 1.9058740990043983]
print(result.mode())
[0.08926614291105403, [1.90000000, 1.90000000, 1.90000000]]
plot_rmle(result)
plot_rmle(result,plt_type='surface')
\end{python}
	
\begin{figure}[htp]
	
	\centering
	\includegraphics[width=.3\textwidth]{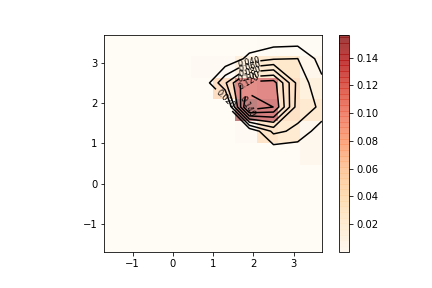}\hfill
	\includegraphics[width=.3\textwidth]{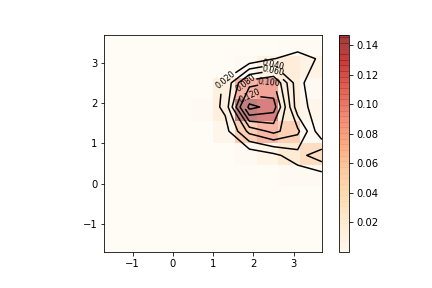}\hfill
	\includegraphics[width=.3\textwidth]{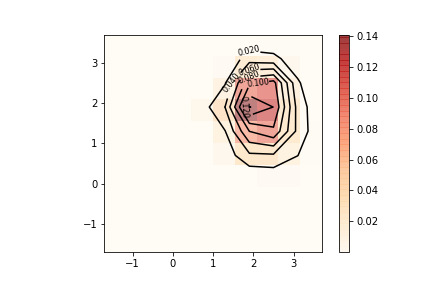}
	
	\caption{Contour plots of joint bivariate marginal distributions of $\hat{f_\bbeta}$: $\hat{f}_{\beta_0,\beta_1}$ (left), $\hat{f}_{\beta_0,\beta_2}$ (middle), and $\hat{f}_{\beta_0,\beta_1}$ (right).}
	\label{fig:3d_10_grid_cont}
	
\end{figure}

\begin{figure}[htp]
	
	\centering
	\includegraphics[width=.3\textwidth]{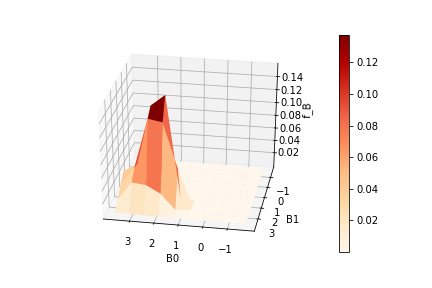}\hfill
	\includegraphics[width=.3\textwidth]{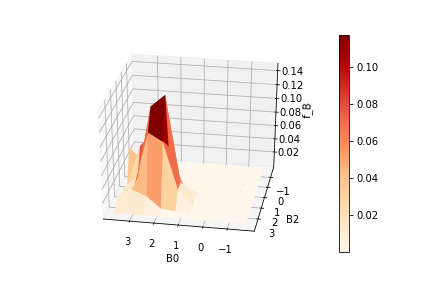}\hfill
	\includegraphics[width=.3\textwidth]{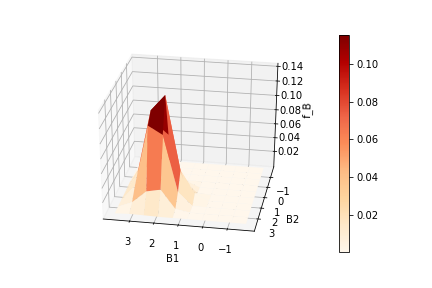}
	
	\caption{Surface plots of joint bivariate marginal distributions of $\hat{f_\bbeta}$: $\hat{f}_{\beta_0,\beta_1}$ (left), $\hat{f}_{\beta_0,\beta_2}$ (middle), and $\hat{f}_{\beta_0,\beta_1}$ (right).}
	\label{fig:3d_10_grid_surface}
	
\end{figure}

In the three dimensional application we use the regularization functional \pythoninline{sobolev_3d} and supply it as an argument to the \pythoninline{rmle()} function along with the transformation matrix and $\alpha$. The results show the effect of the level of discretezation on the estimate. It is possible to achieve more accurate estimates with more grid points in conjunction with a narrower grid range, but with a significantly higher computational cost. 

\medskip

\begin{python}
grid_beta_alt = grid_set(20,3,B0_range=[0,3],\ 
B1_range=[0,3],B2_range=[0,3])
print(grid_beta.numgridpoints())
8000
T2 = transmatrix(sample,grid_beta_alt)
result2 = rmle(sobolev_norm_penal2d,0.3,T2)
print(result2.ev())
[2.102855111361121, 2.077365765266784, 1.9697452677152947]
print(result2.mode()[0])
[[0.2008050879224736, [2.025, 2.025, 2.025]]
plot_rmle(result2)
plot_rmle(result2,plt_type='surface')
\end{python}

\begin{figure}[htp]
	
	\centering
	\includegraphics[width=.3\textwidth]{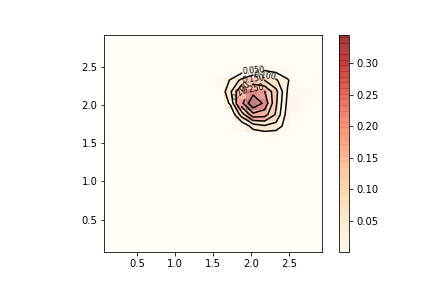}\hfill
	\includegraphics[width=.3\textwidth]{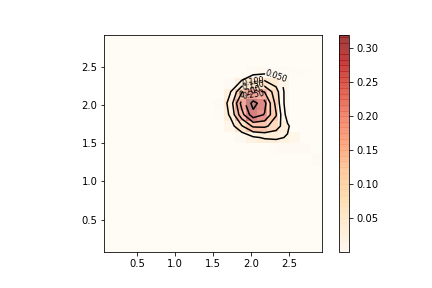}\hfill
	\includegraphics[width=.3\textwidth]{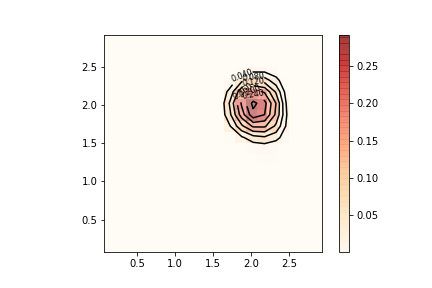}
	
	\caption{Contour plots of joint bivariate marginal distributions of $\hat{f_\bbeta}$: $\hat{f}_{\beta_0,\beta_1}$ (left), $\hat{f}_{\beta_0,\beta_2}$ (middle), and $\hat{f}_{\beta_0,\beta_1}$ (right).}
	\label{fig:3d_20_grid_cont}
	
\end{figure}

\begin{figure}[htp]
	
	\centering
	\includegraphics[width=.3\textwidth]{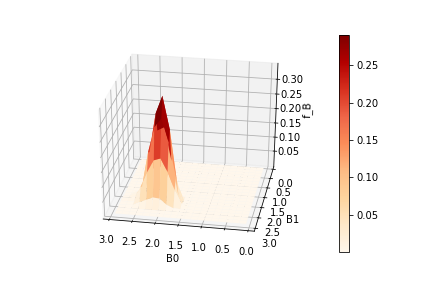}\hfill
	\includegraphics[width=.3\textwidth]{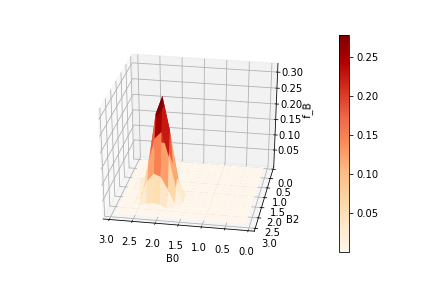}\hfill
	\includegraphics[width=.3\textwidth]{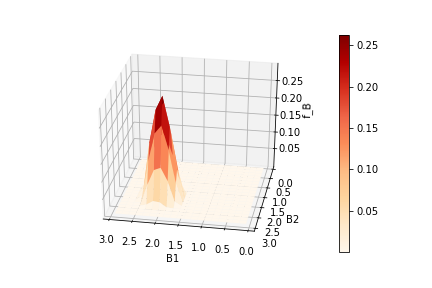}
	
	\caption{Surface plots of joint bivariate marginal distributions of $\hat{f_\bbeta}$: $\hat{f}_{\beta_0,\beta_1}$ (left), $\hat{f}_{\beta_0,\beta_2}$ (middle), and $\hat{f}_{\beta_0,\beta_1}$ (right).}
	\label{fig:3d_20_grid_surface}
	
\end{figure}

In this example, the number of grid points $\hat{f_\bbeta}$ is estimated over is set to 20 on each axis which amounts to a total of 8,000 grid points. The additional level of discretezation due to the increased number of grid points and the smaller grid range provided resulted in an estimate $\hat{f_\bbeta}$.

The next example will demonstrate the case when the underlying joint density being reconstructed has a single mode at the center of the grid. Both methods of dealing with this issue described in section \ref{sec:trans_matrix} will be illustrated.

\medskip

\begin{python}
mu = [[0,0,0],[0,0,0],[0,0,0]]
sample = sim_sample(n = 5000,dim = 3, \
beta_mu = mu)
grid_beta = grid_set(20,3,B0_range=[-1,2],\ 
B1_range=[-1.5,1.5],B2_range=[-1.5,1.5])
T = transmatrix(sample,grid_beta)
result = rmle(sobolev_3d,0.15,T)
print(result.ev())
[-0.19254733276928582, 0.005554898823827626, -0.009788891289943112]
print(result.mode())
.14336637945812933, [-0.024999999999999967, 0.075, -0.075]]
plot_rmle(result)
plot_rmle(result,plt_type='surface')
\end{python}

\begin{figure}[htp]
	
	\centering
	\includegraphics[width=.3\textwidth]{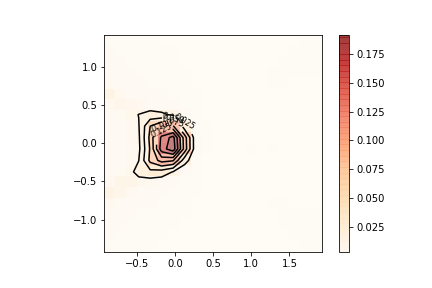}\hfill
	\includegraphics[width=.3\textwidth]{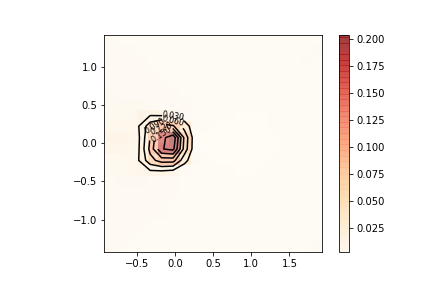}\hfill
	\includegraphics[width=.3\textwidth]{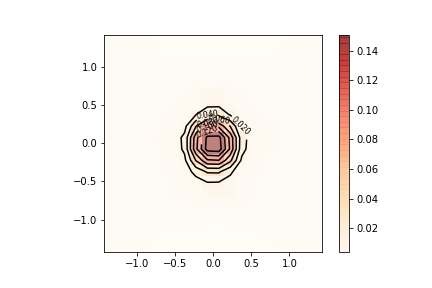}
	
	\caption{Contour plots of joint bivariate marginal distributions of $\hat{f_\bbeta}$: $\hat{f}_{\beta_0,\beta_1}$ (left), $\hat{f}_{\beta_0,\beta_2}$ (middle), and $\hat{f}_{\beta_0,\beta_1}$ (right).}
	\label{fig:3d_20_zero_grid_contour}
	
\end{figure}

\begin{figure}[htp]
	
	\centering
	\includegraphics[width=.3\textwidth]{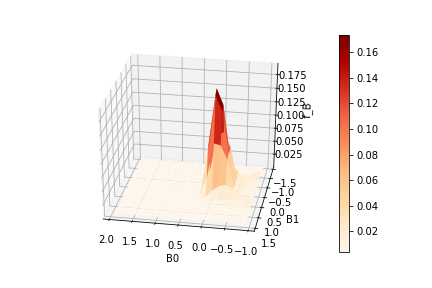}\hfill
	\includegraphics[width=.3\textwidth]{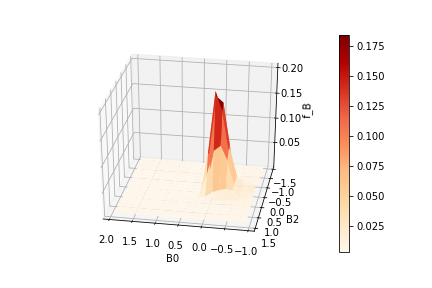}\hfill
	\includegraphics[width=.3\textwidth]{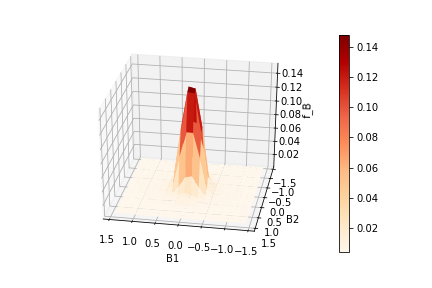}
	
	\caption{Surface plots of joint bivariate marginal distributions of $\hat{f_\bbeta}$: $\hat{f}_{\beta_0,\beta_1}$ (left), $\hat{f}_{\beta_0,\beta_2}$ (middle), and $\hat{f}_{\beta_0,\beta_1}$ (right).}
	\label{fig:3d_20_zero_grid_surface}
	
\end{figure}
The following example demonstrates how to apply the shifting algorithm briefly mentioned in section \ref{sec:trans_matrix}. The algorithm is implemented in python by adding $c \sim \unif(a,b)$ to the intercept, where $a$ and $b$ are determined by the size of the grid. This applies a transformation that can be back-transformed after estimation by adjusting the grid over which $\hat{f_\bbeta}$ is estimated over. This is repeated across ten different shifted samples, then a simple k-means clustering algorithm is applied using \pythoninline{sklearn.cluster.KMeans()} on the $L2$ penalties of \{$\hat{f_\bbeta}_1, \hdots, \hat{f_\bbeta}_{10}$\}. The reconstruction, $\hat{f_\bbeta}_j$, that is closest to the centroid of the largest cluster is then output as the solution.

\begin{python}
grid_beta = grid_set(20,3,B0_range=[-1.5,1.5],\ 
B1_range=[-1.5,1.5],B2_range=[-1.5,1.5])
T = transmatrix(sample,grid_beta)
result_shift = rmle(sobolev_3d,0.15,T,shift=True)
print(result_shift.ev())
[-0.12974928815245057, -0.004493337754614205, -0.0052980597609372385]
print(result_shift.mode())
[[0.14624599807641833, [0.009524681155561987, -0.075, -0.075]]
plot_rmle(result)
plot_rmle(result_shift,plt_type='surface')
\end{python}

\begin{figure}[htp]
	
	\centering
	\includegraphics[width=.3\textwidth]{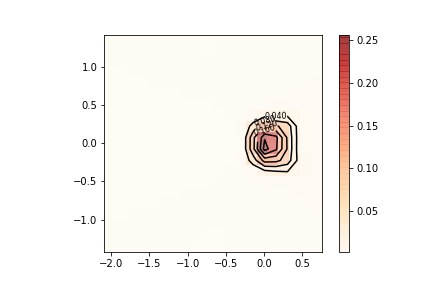}\hfill
	\includegraphics[width=.3\textwidth]{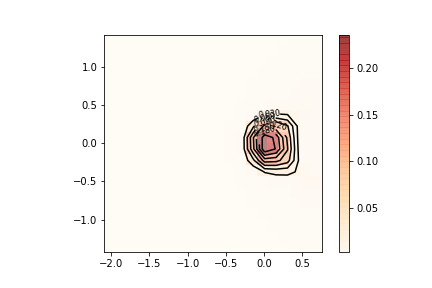}\hfill
	\includegraphics[width=.3\textwidth]{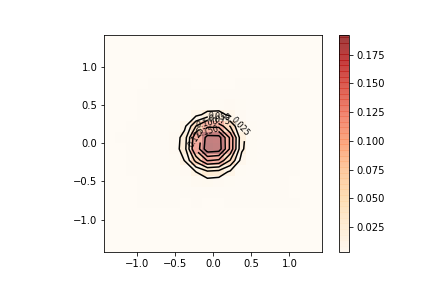}
	
	\caption{Contour plots of joint bivariate marginal distributions of $\hat{f_\bbeta}$: $\hat{f}_{\beta_0,\beta_1}$ (left), $\hat{f}_{\beta_0,\beta_2}$ (middle), and $\hat{f}_{\beta_0,\beta_1}$ (right).}
	\label{fig:3d_20_grid_contour_shift}
	
\end{figure}

\begin{figure}[htp]
	
	\centering
	\includegraphics[width=.3\textwidth]{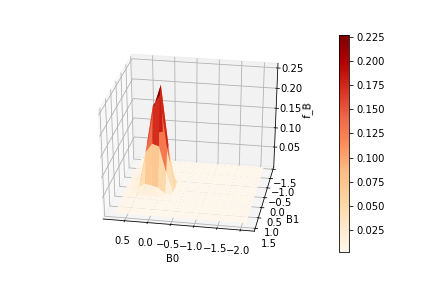}\hfill
	\includegraphics[width=.3\textwidth]{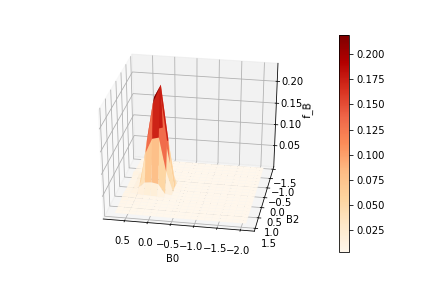}\hfill
	\includegraphics[width=.3\textwidth]{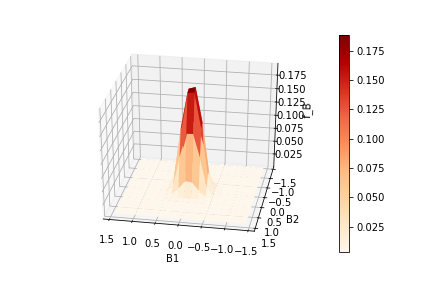}
	
	\caption{Surface plots of joint bivariate marginal distributions of $\hat{f_\bbeta}$: $\hat{f}_{\beta_0,\beta_1}$ (left), $\hat{f}_{\beta_0,\beta_2}$ (middle), and $\hat{f}_{\beta_0,\beta_1}$ (right).}
	\label{fig:3d_20_grid_surface_shift}
	
\end{figure}
The final example demonstrates how to use the module on a real dataset, and shows the user how to load the data into Python from a comma-separated value (CSV) format and how to pre-process it, if necessary, into a usable form. The data used here are from the British Family Expenditure Surey that was used in \cite{Dunker2019,dunker2021regularized}.

The model is identified as follows:
\begin{align*}
	BS_i = \beta_{0,i} + \beta_{1,i}\ln(TotalExpenditure_i) + \beta_{2,i}\ln(FoodPrices_i) + \epsilon_i
\end{align*}
In the case of this dataset the regressors need to undergo a linear transformation before it generating the transformation matrix to be used in the algorithm. This is to ensure that the grid is sufficiently covered by the hyperplanes generated by the sample observations. For more details (refer to the section of our paper discussing this issue).
The OLS estimate for the random coefficients suggests $f_\bbeta$ is centralized at (0.262755,0.0048,-0.00069) for a subsample size of $n = 5000$, which requires the application of one of the two methods described above to circumvent the problem that arises from a mode close to $(0,0,0)$.
\medskip
\begin{python}
data = np.genfromtxt('filename.csv', delimiter=',')
data = data[1:-1,1:4]
data = data[np.random.randint(1,len(data),5000),:]
data = data[~np.isnan(data).any(axis=1)]
ones = np.repeat(1,len(data))
real_data_sample = np.c_[ones, \ 
25*data[:,2]-0.3,data[:,1]-5,data[:,0]]
\end{python}

\medskip

The code block above loads the data from the a csv file, selects a random subsample of 5,000 observations, and applies the linear transformations on the data necessary. 

\medskip
\begin{python}
grid_beta  = grid_set(20,3,B0_range=[-0.75,1.25],\
B1_range=[-1,1],B2_range=[-1,1])
T = transmatrix(real_data_sample,grid_beta)
result = rmle(sobolev_rmle_2d,0.25,T,shift=True)
print(result.ev())
[0.2139255519131858, -0.0022826617746834394, -0.10875379495199464]
print(result.mode()[0:1])
[[0.19533254824055002, [0.3, -0.05, -0.15000000000000002]]
plot_rmle(result)
plot_rmle(result,plt_type='surface')
\end{python}

\begin{figure}[htp]
	
	\centering
	\includegraphics[width=.3\textwidth]{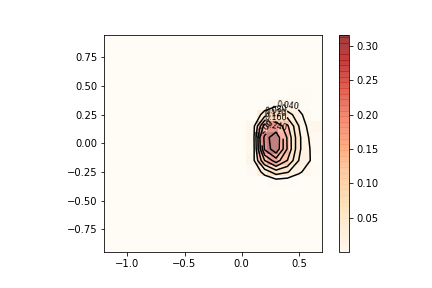}\hfill
	\includegraphics[width=.3\textwidth]{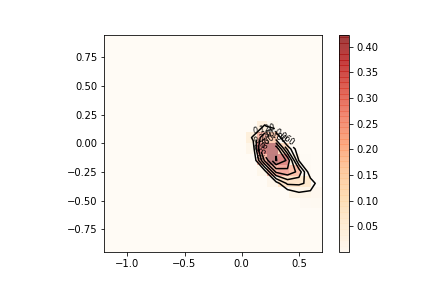}\hfill
	\includegraphics[width=.3\textwidth]{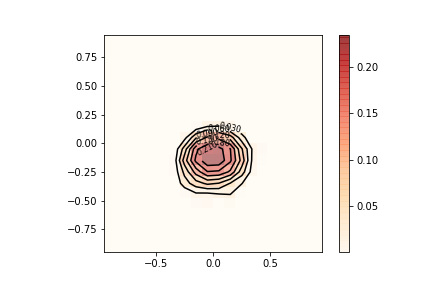}
	
	\caption{Contour plots of joint bivariate marginal distributions of $\hat{f_\bbeta}$: $\hat{f}_{\beta_0,\beta_1}$ (left), $\hat{f}_{\beta_0,\beta_2}$ (middle), and $\hat{f}_{\beta_0,\beta_1}$ (right).}
	\label{fig:3d_20_real_grid_contour}
	
\end{figure}

\begin{figure}[htp]
	
	\centering
	\includegraphics[width=.3\textwidth]{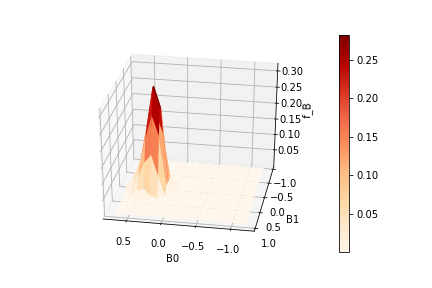}\hfill
	\includegraphics[width=.3\textwidth]{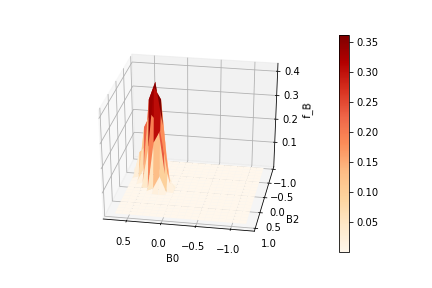}\hfill
	\includegraphics[width=.3\textwidth]{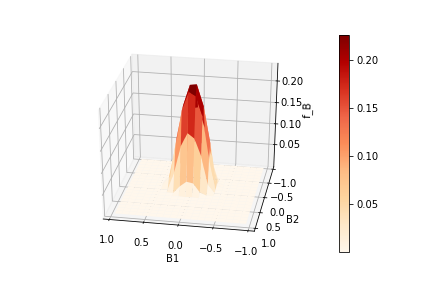}
	
	\caption{Surface plots of joint bivariate marginal distributions of $\hat{f_\bbeta}$: $\hat{f}_{\beta_0,\beta_1}$ (left), $\hat{f}_{\beta_0,\beta_2}$ (middle), and $\hat{f}_{\beta_0,\beta_1}$ (right).}
	\label{fig:3d_20_real_grid_surface}
	
\end{figure}

Included in the module is an option to fit a spline on the estimate $\hat{f_\beta}$. This would be the most computationally feasible option if the user needs a finer grid. The \pythoninline{spline_fit()} function takes two essential arguments. The first positional argument is the \pythoninline{class RMLEResult} object, and second is the number of grid points on each axis.

\begin{python}
spline = spline_fit(result,num_grid_points = 400)
plot_rmle(spline)
plot_rmle(spline,plt_type='surface')
\end{python}

\begin{figure}[htp]
	
	\centering
	\includegraphics[width=.3\textwidth]{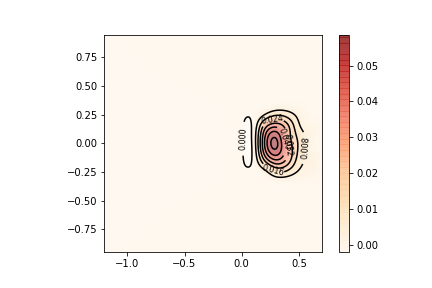}\hfill
	\includegraphics[width=.3\textwidth]{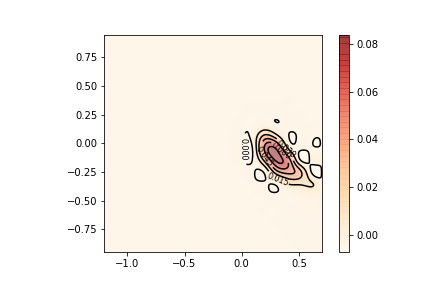}\hfill
	\includegraphics[width=.3\textwidth]{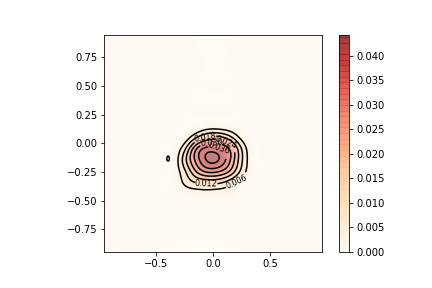}
	
	\caption{Contour plots of the spline interpolated joint bivariate marginal distributions of $\hat{f_\bbeta}$: $\hat{f}_{\beta_0,\beta_1}$ (left), $\hat{f}_{\beta_0,\beta_2}$ (middle), and $\hat{f}_{\beta_0,\beta_1}$ (right).}
	\label{fig:3d_20_real_grid_contour_spline}
	
\end{figure}

\begin{figure}[htp]
	
	\centering
	\includegraphics[width=.3\textwidth]{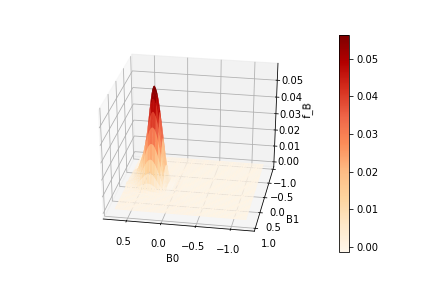}\hfill
	\includegraphics[width=.3\textwidth]{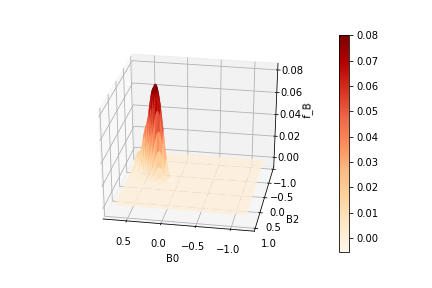}\hfill
	\includegraphics[width=.3\textwidth]{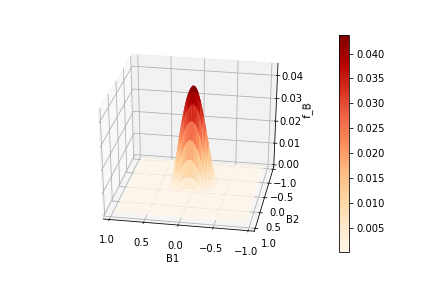}
	
	\caption{Surface plots of the spline interpolated joint bivariate marginal distributions of $\hat{f_\bbeta}$: $\hat{f}_{\beta_0,\beta_1}$ (left), $\hat{f}_{\beta_0,\beta_2}$ (middle), and $\hat{f}_{\beta_0,\beta_1}$ (right).}
	\label{fig:3d_20_real_grid_surface_spline}
	
\end{figure}

\bibliographystyle{apalike}
\bibliography{PyRMLE_Paper}

\end{document}